\documentclass[5p,twocolumn,times,number]{elsarticle}

\usepackage{graphicx}
\usepackage{amsmath}   
\usepackage{lineno}

\begin{document}

\begin{frontmatter}

\title{Design of a SiPM-based cluster\\ for the Large Size Telescope camera of the Cherenkov Telescope Array}

\author[add1]{M.~Mallamaci\corref{cor}}
\ead{manuela.mallamaci@pd.infn.it}
\author[add1]{B.~Baibussinov} 
\author[add1,add2]{G.~Busetto}
\author[add1]{D.~Corti}
\author[add1,add3,add4,add5]{A.~De~Angelis} 
\author[add6]{F.~Di~Pierro}
\author[add1,add2]{M.~Doro}
\author[add3]{L.~Lessio}
\author[add1,add2]{M.~Mariotti}
\author[add1,add2]{R.~Rando}
\author[add1,add2]{E.~Prandini}
\author[add6,add7]{P.~Vallania}
\author[add6,add8]{C.~F.~Vigorito}
\author{for the CTA LST project}

\cortext[cor]{Corresponding author}

\address[add1]{Istituto Nazionale di Fisica Nucleare, Sezione di Padova, I-35131 Padova, Italy}
\address[add2]{Dipartimento di Fisica e Astronomia \emph{``G. Galilei''}, Universit\`a di Padova, I-35131 Padova, Italy}
\address[add3]{Istituto Nazionale di Astrofisica - Osservatorio Astronomico di Padova, I-35122, Padova, Italy}
\address[add4]{Dipartimento di Matematica, Informatica e Fisica, Universit\`a di Udine, I-33100 Udine, Italy}
\address[add5]{Laboratorio de Instrumenta\c{c}ao e Particulas and Instituto Superior Tecnico, Lisboa, Portugal}
\address[add6]{Istituto Nazionale di Fisica Nucleare, Sezione di Torino, I-10125 Torino, Italy}
\address[add7]{Istituto Nazionale di Astrofisica - Osservatorio Astrofisico di Torino, I-10025 Torino, Italy}
\address[add8]{Dipartimento di Fisica, Universit\`a degli Studi di Torino, I-10125 Torino, Italy}

\begin{abstract}
A Silicon Photomultiplier (SiPM)-based photodetector is being built to demonstrate
its feasibility for an alternative silicon-based camera design for the
Large Size Telescope (LST) of the Cherenkov
Telescope Array. It has been designed to match the size of the
standard Photomultiplier Tube (PMT) cluster unit and to be compatible with
mechanics, electronics and focal plane optics of the first LST camera.
Here, we describe the overall SiPM cluster design along with the main
differences with respect to the currently used PMT cluster unit. The fast
electronics of the SiPM pixel and its layout are also presented. In order
to derive the best working condition for the final unit, we measured the
SiPM performances in terms of gain, photo-detection efficiency and
cross-talk. One pixel, a unit of 14 SiPMs, has been built. We will discuss
also some preliminary results regarding this device and we will highlight
the future steps of this project.
\end{abstract}


\end{frontmatter}
\section{Introduction}
In the last decade, very high energy gamma-ray astronomy has enormously improved. 
A wealth of information has been collected by H.E.S.S., MAGIC and VERITAS, the current generation
 of Imaging Air Cherenkov Telescopes (IACTs), providing a complex and fascinating picture of the gamma-ray Universe at the highest energies. 
The Cherenkov Telescope Array (CTA) represents the future for IACTs \cite{cta}. It will provide an improvement in sensitivity by at least an order of 
magnitude as compared to current IACTs and it will cover a very large range in energy, going from tens of GeV up to  about 100 TeV. \\
Three types of telescopes are foreseen, depending on the energy range to detect: 23 m - diameter Large-Sized Telescopes (LSTs), 
12 m - diameter Medium-Sized Telescopes (MSTs) and 4 m - diameter Small-Sized Telescopes (SSTs).  The first LST is under construction on the Canary
 Island of La Palma, the site chosen for the Northern array.  It is foreseen also the construction of a second array in the Southern hemisphere \cite{cta}.\\
 LSTs of CTA are designed to enhance the sensitivity below 200~-~300 GeV and to lower the effective threshold down to~20~-~30~GeV. The science case of these 
 instruments is described in \cite{wbCTA}.
Since LST is designed to cover the lower energy range of CTA,  it must collect the maximum number of Cherenkov photons from gamma-induced air showers, 
consisting in weak light flashes lasting few ns. The reflective surface area and the telescope size itself have to be maximized:
 it is important that the efficiency of the photo-sensor is as large as possible to capitalize on the high cost of the mechanics. 
 The baseline design of LST includes a focal-plane camera based on photo multiplier tubes (PMTs), with a  field of view of about 4.5$^\circ$.  
It comprises 265 PMT modules. Each module, called \textit{cluster}, has 7 channels, providing the camera with a total of 1855 channels. 
  Hamamatsu PMTs with a peak quantum efficiency of $\sim 42$\% (R11920-100) are used \cite{LSTcamera}.
\section{Design of a SiPM-cluster for the LST camera}
\begin{figure}
\centering
\includegraphics[width=0.46\linewidth]{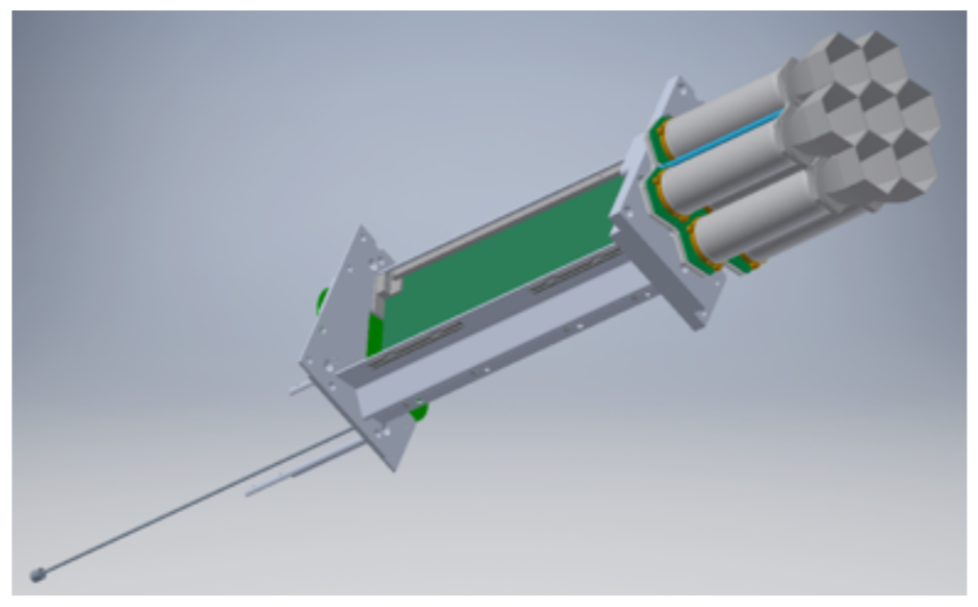}
\includegraphics[width=0.4\linewidth]{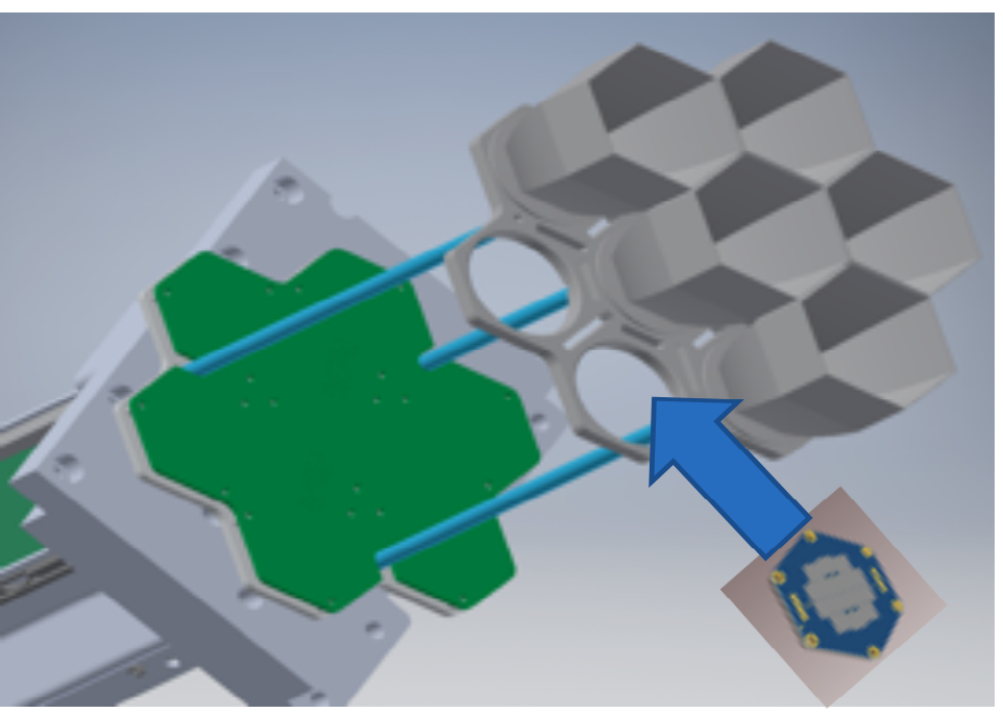}
\includegraphics[width=0.43\linewidth]{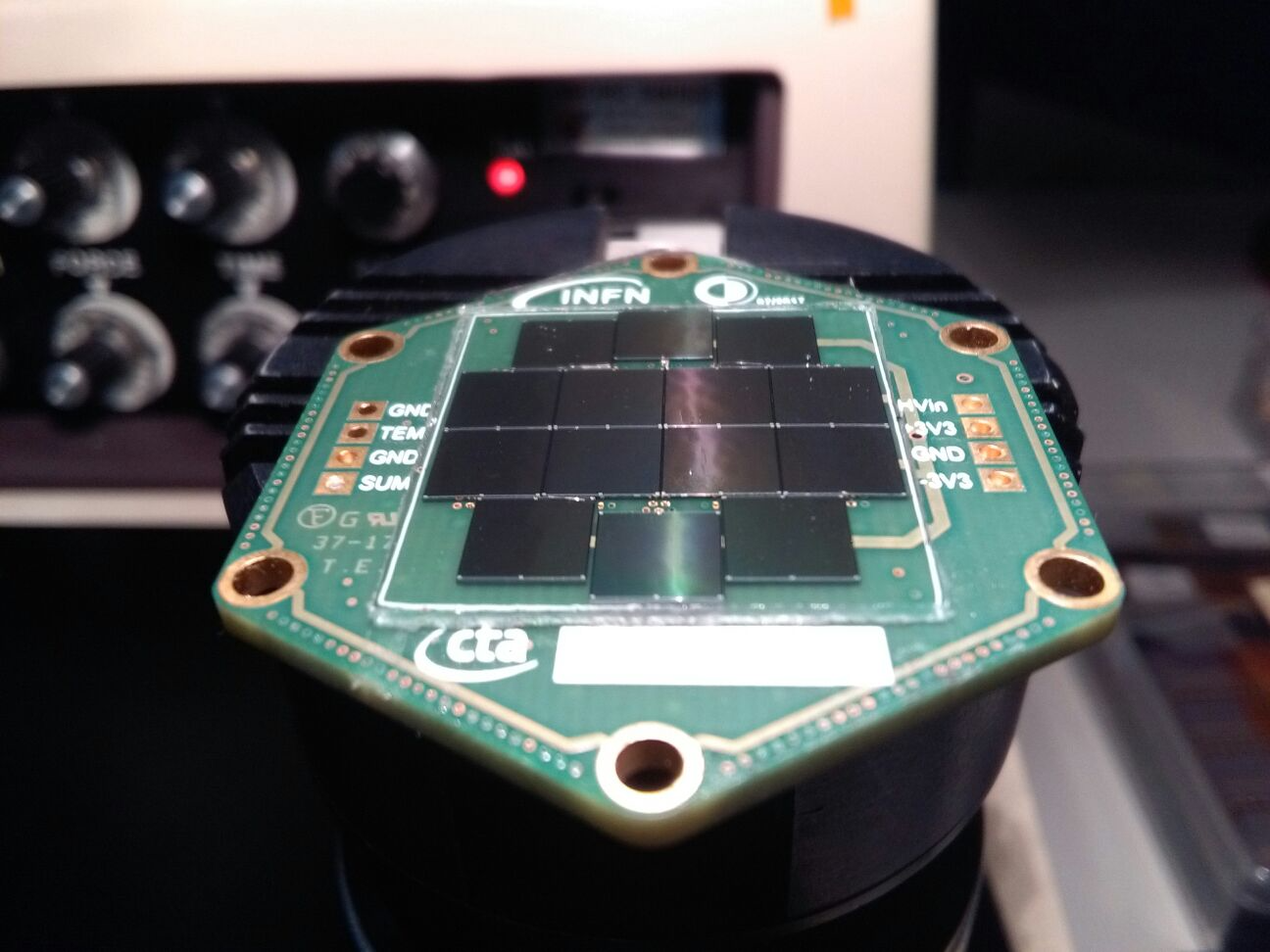}
\includegraphics[width=0.43\linewidth]{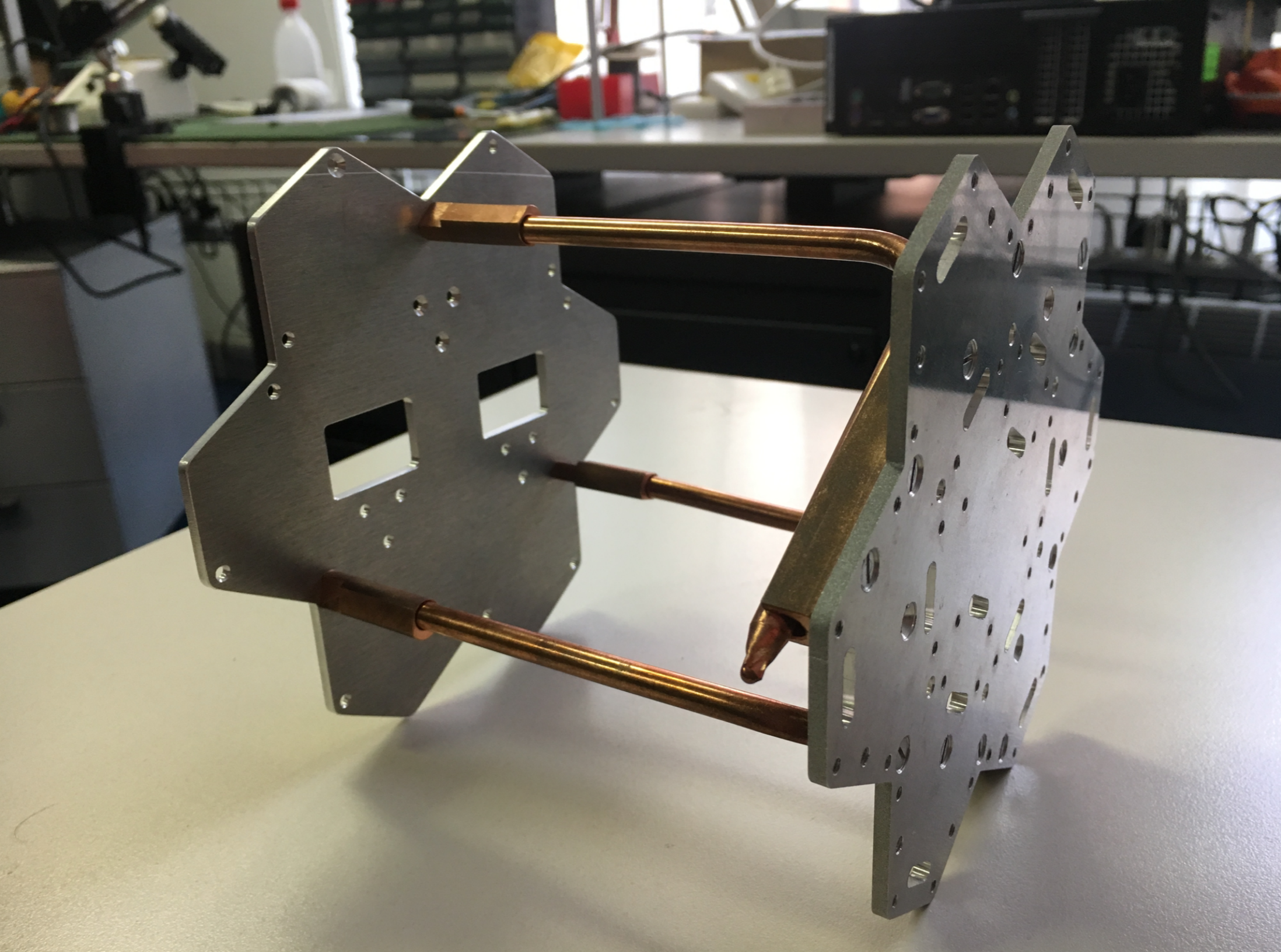}
\caption{Top - Mechanical structure of a LST PMT cluster. Bottom - Mechanical structure of a SiPM cluster.  }
\label{fig:cluster}
\end{figure}

The goal of this research is to develop a Silicon Photomultiplier (SiPM)-based prototype cluster for LST, as an alternative camera design for the Southern site. 
This will be the solid-state equivalent of a PMT, having a few cm$^2$ of sensitive area, high photon detection efficiency (PDE), 
good single-photon sensitivity, and time response around 2-3 ns. In general, SiPM detectors are an almost 
ideal device to be used in the focal plane of a Cherenkov telescope, having a high detection efficiency and allowing a higher duty cycle. 
SiPM technology is already successfully exploited for the First G-APD Cherenkov Telescope (FACT), operating for more than 6 years \cite{Biland}, and for SSTs of CTA (see e.g. \cite{montaruli}), 
and development of a silicon camera prototype for the LSTs has been ongoing \cite{oldcluster}.
Fig. \ref{fig:cluster} (top-left) shows the mechanical structure design of a PMT LST cluster. The pre-amplifier circuits 
for the anode signal and the resistors for the dynode voltage divider lie on an aluminium plate, cooled at a temperature of 15$^\circ$C. 
Fig. \ref{fig:cluster} (top-right) reports the mechanical scheme of the proposed SiPM cluster. The electronics and the sampling system are the same, whereas the
 high voltage for PTMs is replaced by a low voltage (and low noise) power supply for the SiPMs, with an interface board.
In order to adapt the SiPMs pixel layout to the current camera design, each sensor will cover 0.1$^\circ$ and match the exit window 
of the light concentrator with hexagonal entrance pupil \cite{newLG}.
\section{SiPM characterization}
\begin{figure}
\centering
\includegraphics[width=0.99\linewidth]{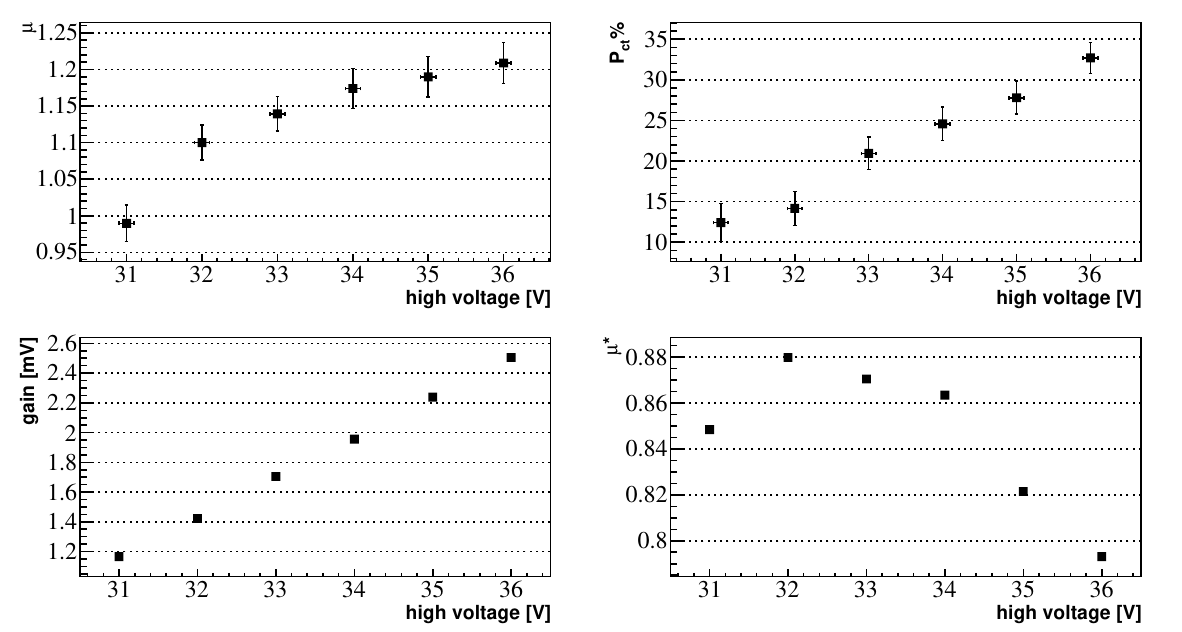}
\caption{Results on FBK NUV HD3\_2 SiPM (single photo-sensor), obtained for $\lambda$=376 nm. See text for details.}
\label{fig:results}
\end{figure}

In this work,  we have characterized near-UV FBK (\textit{Fondazione Bruno Kessler})
SiPMs (``HD3\_2'') with dimension of 6$\times$6 mm$^2$. The PDE of these sensors is
 60\% at 350 nm. 
The single sensor has been connected to an amplifier powered by a voltage of -3.3 V and 3.3 V. This amplifier has 2 ns of full width at half maximum
(FWHM) and it preserves a good signal-to-noise ratio (S/N).  Results of the analysis are shown in Fig. \ref{fig:results} as a function of the high voltage. 
They have been obtained by illuminating the sensors with a light-emitting diode (LED) source at $\lambda$=376 nm (Picoquant PLS8-2-592). A large number (10000) 
of waveforms has been collected.
From the peak voltage distribution, we determined the number of collected photoelectrons (indicated as $\mu$), the cross-talk probability (P$_{ct}$\%), 
the gain and S/N. The latter is  reported as $\mu$* = (S/N)$^2$. In the hypothesis of a Poissonian distribution,
 $\mu$* and $\mu$ would be the same, in realistic condition they are not because of the cross-talk effect.
One can see that the gain changes linearly with the applied voltage. In addition, (S/N) is maximized at $\sim$ 32 V, where the cross-talk is 14.2$\pm$2.1(stat)\%. 
 According to \cite{vinogradov}, the excess noise factor due to stochastic noise of secondary events can be defined as ENF~=~1~+~$P_{ct}$.
 Its value is $\sim$1.14 at the optimal working point.  
\section{SiPM pixel characterization and future studies}
This project involves the production of 7 pixels. The pixel design  
consists of 14 sensors. Each sub-element is connected to a simple passive high-pass filter, to differentiate the signal and to decouple 
the capacitance of the element from the other elements, thus keeping the pulse rise-time around 1 ns. The signals from all the elements 
are summed up in an adder circuit and then propagated to the acquisition electronics.
Following this design, we built and characterized two pixels with 14 SiPMs of the model described above (one is shown in Fig. \ref{fig:cluster}, bottom-left).
The pixel behaves as a sum of 14 objects,
  within the errors, and  preserves a FWHM of less than 2.7 ns. 
 Globally, its performance is compatible with the operational requirements: the electronic noise is 0.78 mV and the dynamical range is around 1000 (defined as the ratio A*/$\sigma_e$, with A* amplitude of the signal before saturation and $\sigma_e$, electronic noise).\\
 One of the next steps for this project is to design an optical system. The goal is to map the focal plane into the pixelated camera, avoiding dead areas between pixels and to compress the hexagonal entrance pupil area into the sensor area. In addition to the optical system, we will test how to drive the heat from the power control board to the cooling plate, which is 15~cm below. For this purpose a set of heat pipes will be applied. A prototype of the mechanical structure built for this project is reported in Fig. \ref{fig:cluster} (bottom-right).
\section*{Acknowledgements}
We gratefully acknowledge financial support from the agencies and organisations listed here: https://www.cta-observatory.org/consortium\_acknowledgments/

\end{document}